\documentclass[5p,times,sort&compress]{elsarticle}
\usepackage{amssymb}
\usepackage{color}
\usepackage{natbib}
\usepackage[dvipsnames]{xcolor}
\colorlet{orange}{orange!120!}
\usepackage{graphicx}
\usepackage{dcolumn}
\usepackage{bm}
\usepackage{amstext}
\usepackage{verbatim}
\usepackage[colorlinks,citecolor=blue,linkcolor=blue,anchorcolor=blue,filecolor=blue,urlcolor=blue]{
hyperref}
\usepackage{tabularx}
\usepackage{dcolumn} 
\usepackage{ulem} 
\usepackage{longtable}
\usepackage{soul}
\newcolumntype{d}[1]{D{.}{.}{#1}} 
\definecolor{americanrose}{rgb}{1.0, 0.01, 0.24}

\journal{arxiv}

\begin{document}
\begin{frontmatter}
\title{Short-range correlation effects on the neutron star cooling}

\author{Lucas A. Souza$^1$}
\author{Rodrigo Negreiros$^2$}
\author{Mariana Dutra$^1$}
\author{D\'ebora P. Menezes$^3$}
\author{Odilon Louren\c{c}o$^1$}
\address{
\mbox{$^1$Departamento de F\'isica, Instituto Tecnol\'ogico de Aeron\'autica, DCTA, 12228-900, 
S\~ao Jos\'e dos Campos, SP, Brazil}\\
\mbox{$^2$ Instituto de F\'isica, Universidade Federal Fluminense 20420, Niter\'oi, RJ, Brazil}
\mbox{$^3$ Departamento de F\'isica, Universidade Federal de Santa Catarina, Florian\'opolis, SC, 
CP 476, CEP 88.040-900, Brazil}\\
}

\begin{abstract} 
Short range correlations (SRC) have been known to be an important aspect of nuclear theory for some 
time. Recent works have re-ignited interest on this topic, particularly due to the fact that it has 
recently been demonstrated that SRC may be responsible for breaking pairing gaps in nuclear matter. 
In this work we revisit the concept of SRC for beta equilibrated matter in neutron stars. We 
construct two equivalent models, with and without SRC and proceed to investigate the thermal 
evolution of stars described by such models. We show that SRC play a major role in the thermal 
evolution of neutron stars. It will be shown that while the SRC largely leaves the macroscopic 
properties of the star unaltered, it significantly alters the proton fraction, thus leading to an 
early onset of the direct Urca (DU) process, which in turns leads to stars exhibiting much faster 
cooling.
\end{abstract}
\end{frontmatter}

\section{Introduction}

Claims that the pairing interaction plays an important role in astrophysics are very common. In 
fact, superfluidity has been investigated both in infinite nuclear matter and finite nuclei 
\cite{dean2003} and it is a key ingredient in the neutron star cooling process  
\cite{Shternin2011,Yakovlev2011,Page2011a,2004ApJS..155..623P,Yakovlev2004,Negreiros2011,
Negreiros2012,Beloin2016,Negreiros2017}. On the other hand, short-range correlations (SRC) have 
also become a hot topic recently \cite{nature,hen2017} and one of the reasons for the new enthusiasm 
on an old topic is the fact that SRC can be responsible for breaking the pairs around the Fermi 
surface and hence, reduce the pairing gap mechanism \cite{arnaud2016,arnaud2017}. 

One of the experiments used to probe the existence of the short-range correlations took place in 
the Thomas Jefferson National Accelerator Facility (JLab), and it was concluded that the high 
energy incident beams of protons, or electrons, in the $^{12}\rm C$ nucleus, give rise to ejected 
nucleon pairs~\cite{science1} submitted to the short-range correlations 
(SRC)~\cite{nature,hen2017,ye2018,Egiyan2006,Frankfurt1993,Fomin2012,Atti2015,Shneor2007,Tang2003,
Li2019,Schmookler2019,Duer2019,Ryck2019,Chen2017}. It was found that about $20\%$ of the nucleons in 
the $^{12}\rm C$ nucleus form such correlated structures. Among them, $90\%$ are composed by the 
$np$ pair with the remaining $nn$ and $pp$ pairs divided into $5\%$ for each one of them. The 
dominance of the correlated $np$ pair is also observed in experiments involving $^{27}\rm Al$, 
$^{56}\rm Fe$ and $^{208}\rm Pb$~ nuclei~\cite{science2}.

Effects of short range correlations on infinite nuclear matter equations of state (EOS) are known 
not to be negligible for a long time \cite{prafulla}. The very same kind of EOS, obtained from 
relativistic models, are often used to describe neutron star matter and as input to the analyses of 
the star cooling process.  

In the present work, we explicitly include SRC in a relativistic mean field model and later analyze 
the effects on the thermal evolution of neutron stars. For this purpose we proceed to calculate the 
macroscopic properties of stars in two equivalent models, with and without SRC. The family of stars 
constructed is then used in thermal evolution calculations. As we will see SRC play a major role in 
the thermal history of neutron stars. As of the writing of this letter, and to the best of or 
knowledge this is the first study of this nature.

\section{Short-range correlations in a relativistic mean-field model}

From the theoretical point of view, a direct effect of the SRC in the nucleon dynamics is the change 
in its momentum distribution function, $n(k)$. For our study, we closely follow the structure for 
$n(k)$ proposed in Ref.~\cite{Cai2016}, namely, $n_{n,p}(k_F, y) = \Delta_{n,p}$ for 
$0<k<k_{F\,{n,p}}$, and $n_{n,p}(k_F, y) = C_{n,p}(k_{F\,{n,p}}/k)^4$ for $k_{F\,{n,p}}<k<\phi_{n,p} 
k_{F\,{n,p}}$. The proton fraction is given by $y=\rho_p/\rho$, with $\rho_p$ being the proton 
density and $\rho=2k_F^3/(3\pi^2)$ the total one. The $\Delta_{n,p}$ constants, namely, the 
depletion of the Fermi sphere at zero momentum \cite{Czyz1960,Belkov1961,Sartor1980}, are written in 
terms of $C_{n,p}$ and $\phi_{n,p}$ as $\Delta_{n,p}=1 - 3C_{n,p}(1-1/\phi_{n,p})$, where $C_p=C_0[1 
- C_1(1-2y)]$, $C_n=C_0[1 + C_1(1-2y)]$, $\phi_p=\phi_0[1 - \phi_1(1-2y)]$ and $\phi_n=\phi_0[1 + 
\phi_1(1-2y)]$. Analysis of $d(e,e',p)$ reactions and medium-energy photonuclear absorptions allowed 
to determine the value of $C_0=0.161$~\cite{baoanli}. Constraints from nuclear many-body theories 
and the equations of state of cold atoms under unitary condition were used to find $C_n$ in pure 
neutron matter equal to $0.12$, consequently resulting in $C_1=-0.25$~\cite{baoanli}. Furthermore, 
$\phi_0 = 2.38$ was found from systematic investigation of $(e, e')$ reactions and data from 
two-nucleon knockout reactions, as also described in Ref.~\cite{baoanli}. As in pure neutron matter 
$\phi_n$ was found to be $1.04$, it was also possible to determine $\phi_1=-0.56$~\cite{baoanli}.

The structure given by $n_{n,p}$ is used in the equations of state obtained from the following 
Lagrangian density~\cite{Chen2007,Cai2012,Dutra2014,rmfdef,Cai2016},
\begin{eqnarray}
\hspace{-1.5cm}
\mathcal{L} &=& \bar{\psi}\left(i \gamma^{\mu} \partial_{\mu}-M_{\mbox{\tiny nuc}}\right) 
\psi+g_{\sigma} \sigma \bar{\psi} \psi-g_{\omega} \bar{\psi} \gamma^{\mu} \omega_{\mu} \psi  
\nonumber\\ 
&-&\frac{g_{\rho}}{2} \bar{\psi} \gamma^{\mu} \vec{\rho}_{\mu} \vec{\tau} \psi
+\frac{1}{2}\left(\partial^{\mu} \sigma \partial_{\mu} \sigma-m_{\sigma}^{2} 
\sigma^{2}\right)-\frac{A}{3} \sigma^{3}-\frac{B}{4} \sigma^{4} \nonumber\\
&-&\frac{1}{4} F^{\mu \nu} F_{\mu \nu}+\frac{1}{2} m_{\omega}^{2} \omega_{\mu} 
\omega^{\mu}+\frac{C}{4}\left(g_{\omega}^{2} \omega_{\mu} \omega^{\mu}\right)^{2} -\frac{1}{4} 
\vec{B}^{\mu \nu} \vec{B}_{\mu \nu}\nonumber\\
&+&\frac{1}{2} m_{\rho}^{2} \vec{\rho}_{\mu} \vec{\rho}^{\ \mu}
+\frac{1}{2} \alpha_3' g_{\omega}^{2} g_{\rho}^{2} \omega_{\mu} \omega^{\mu} \vec{\rho}_{\mu} 
\vec{\rho}^{\mu},
\label{lagrangian} 
\end{eqnarray}
in which $F_{\mu \nu}=\partial_{\mu} \omega_{\nu}-\partial_{\nu} \omega_{\mu}$ and $\vec{B}_{\mu 
\nu}=\partial_{\mu} \vec{\rho}_{\nu}-\partial_{\nu} \vec{\rho}_{\mu}$. Here, $\psi$ is the nucleon 
field and $\sigma$, $\omega_\mu$ and $\vec{\rho}_{\mu}$ represents the fields of the mesons 
$\sigma$, $\omega$ and $\rho$, respectively. We also used the standard masses of $M_{\mbox{\tiny 
nuc}}=939$~MeV (nucleon rest mass),  $m_\omega = 782.5$~MeV, $m_\rho = 763$~MeV and 
$m_\sigma=500$~MeV. By using mean-field approximation in order to solve the field equations of 
motion we find $m_{\sigma}^{2} \sigma=  g_{\sigma} \rho_{s}-A \sigma^{2}-B \sigma^{3}$, 
$m_{\omega}^{2} \omega_{0}= g_{\omega} \rho-C g_{\omega}\left(g_{\omega} \omega_{0}\right)^{3} 
-\alpha'_3 g_{\omega}^{2} g_{\rho}^{2} \bar{\rho}_{0(3)}^{2} \omega_{0}$, $m_{\rho}^{2} 
\bar{\rho}_{0(3)}= g_\rho\rho_3/2- \alpha'_3g_{\omega}^{2} g_{\rho}^{2} \bar{\rho}_{0(3)} 
\omega_{0}^{2}$ and $[\gamma^\mu (i\partial_\mu - V_\tau ) - M^*)]\psi = 0$, where it is used the 
expectation values of the fields, namely, $\sigma$, $\omega_0$ (zero component) and 
$\bar{\rho}_{0_{(3)}}$ (isospin space third component). Furthermore, one has $V_{\tau} = 
g_\omega\omega_0 + g_\rho\bar{\rho}_{0(3)}\tau_3/2$, 
\mbox{$\rho=\langle\bar{\psi}\gamma^0\psi\rangle$}$=\rho_n+\rho_p$, 
\mbox{$\rho_3=\langle\bar{\psi}\gamma^0\tau_3\psi\rangle$}$=\rho_{p}-\rho_{n}=(2 y-1)\rho$, and 
\mbox{$\rho_s=\langle\bar{\psi}\psi\rangle$}$={\rho_s}_p + {\rho_s}_n$, with
\begin{eqnarray}
{\rho_s}_{n,p} &=& 
\frac{\gamma M^*\Delta_{n,p}}{2\pi^2} \int_0^{k_{F\,{n,p}}}  
\frac{k^2dk}{\left({k^{2}+M^{*2}}\right)^{1/2}} 
\nonumber\\
&+& \frac{\gamma M^*C_{n,p}}{2\pi^2} \int_{k_{F\,{n,p}}}^{\phi_{n,p} {k_{F\,{n,p}}}} 
\left(\frac{{k_F}_{n,p}}{k}\right)^{4}  \frac{k^2dk}{\left({k^{2}+M^{*2}}\right)^{1/2}}.
\label{rhos}
\end{eqnarray}
The effective nucleon mass is defined by $M^*= M_{\mbox{\tiny nuc}}-g_\sigma\sigma$, the degeneracy 
factor is $\gamma=2$ for asymmetric matter, and \mbox{$\tau_3=+1(-1)$} for protons (neutrons).

The energy-momentum tensor $T_{\mu \nu}$, calculated from the Lagrangian density in 
Eq.~(\ref{lagrangian}), yields the energy density and the pressure of the asymmetric system, since 
$\epsilon=\langle T_{00}\rangle$ and $P=\langle T_{ii}\rangle /3$. These quantities are given, 
respectively, by
\begin{eqnarray} 
\epsilon &=& \frac{1}{2} m_{\sigma}^{2} \sigma^{2}+\frac{A}{3} \sigma^{3}+\frac{B}{4} 
\sigma^{4}-\frac{1}{2} m_{\omega}^{2} \omega_{0}^{2}-\frac{C}{4}\left(g_{\omega}^{2} 
\omega_{0}^{2}\right)^{2}
\nonumber\\
&-& \frac{1}{2} m_{\rho}^{2} \bar{\rho}_{0(3)}^{2}+g_{\omega} \omega_{0} \rho+\frac{g_{\rho}}{2} 
\bar{\rho}_{0(3)} \rho_{3} 
  -\frac{1}{2} \alpha'_3 g_{\omega}^{2} g_{\rho}^{2} \omega_{0}^{2} \bar{\rho}_{0(3)}^{2}
\nonumber\\
&+& \epsilon_{\mathrm{kin}}^{p}+\epsilon_{\mathrm{kin}}^{n} 
\label{denerg}
\end{eqnarray}
and
\begin{eqnarray}
p &=&-\frac{1}{2} m_{\sigma}^{2} \sigma^{2}-\frac{A}{3} \sigma^{3}-\frac{B}{4} 
\sigma^{4}+\frac{1}{2} m_{\omega}^{2} \omega_{0}^{2}
+\frac{C}{4}\left(g_{\omega}^{2} \omega_{0}^{2}\right)^{2} 
\nonumber\\
&+&\frac{1}{2} m_{\rho}^{2} \bar{\rho}_{0(3)}^{2}
+\frac{1}{2} \alpha'_3 g_{\omega}^{2} g_{\rho}^{2} \omega_{0}^{2} \bar{\rho}_{0(3)}^{2}  
+p_{\mathrm{kin}}^{p}+p_{\mathrm{kin}}^{n},
\label{press}
\end{eqnarray}
with the following kinetic contributions:
\begin{eqnarray} 
\epsilon_{\text {kin}}^{n,p} &=& \frac{\gamma \Delta_{n,p}}{2\pi^2} \int_0^{{k_{F\,{n,p}}}} 
k^2dk({k^{2}+M^{* 2}})^{1/2}
\nonumber\\
&+& \frac{\gamma C_{n,p}}{2\pi^2} \int_{k_{F\,{n,p}}}^{\phi_{n,p} {k_{F\,{n,p}}}} \left(\frac{{k _F}_{n,p}}{k}\right)^{4}  k^2dk({k^{2}+M^{* 2}})^{1/2},
\end{eqnarray}
and
\begin{eqnarray} 
 p_{\text {kin }}^{n,p} &=&  
\frac{\gamma \Delta_{n,p}}{6\pi^2} \int_0^{k_{F\,{n,p}}}  
\frac{k^4dk}{\left({k^{2}+M^{*2}}\right)^{1/2}} 
\nonumber\\
&+& \frac{\gamma C_{n,p}}{6\pi^2} \int_{k_{F\,{n,p}}}^{\phi_{n,p} {k_{F\,{n,p}}}} \left(\frac{{k 
_F}_{n,p}}{k}\right)^{4}  \frac{k^4dk}{\left({k^{2}+M^{*2}}\right)^{1/2}}.
\end{eqnarray}
The chemical potentials for protons and neutrons are obtained from the general definition 
$\mu_{n,p}=  {\partial \epsilon}/{\partial \rho_{n,p}}$, giving rise to $\mu_{n,p} = 
\mu^{n,p}_{\mathrm{kin (SRC)}}+\Delta_{n,p}\mu^{n,p}_\mathrm{kin}+g_{\omega} \omega_{0} 
\pm g_\rho\bar{\rho}_{0_{(3)}}/2$, with the sign $+$ ($-$) for protons (neutrons). Here, $\mu^{n,p}_{\mathrm{kin}}=({k^2_F}_{n,p}+M^{*2})^{1/2}$ and
\begin{eqnarray} 
\mu^{n,p}_{\mathrm{kin\,(SRC)}} = 3C_{n,p} \left[ \mu^{n,p}_{\mathrm{kin}}
- \frac{\left({\phi_{n,p}^2 {k^2_F}_{n,p} + M^{*2}}\right)^{1/2}}{\phi_{n,p}} \right]
\nonumber\\
+ {4}C_{n,p} {k_F}_{n,p} \ln\left[\frac{\phi_{n,p} {k_F}_{n,p} + 
\left(\phi_{n,p}^2{k_F^2}_{n,p}+M^{*2}\right)^{1/2} }{ {k_F}_{n,p} + \left( {k^2_F}_{n,p} + M^{* 
2}\right)^{1/2}}\right].
\end{eqnarray} 

Since the thermodynamics of the system is completely defined, it is possible to determine the 
coupling constants of the model, namely, $g_\sigma$, $g_\omega$, $g_\rho$, $A$, $B$, $C$ and 
$\alpha'_3$. In order to proceed with this calculation, we use the following bulk parameters: 
$\rho_0=0.15$~fm$^{-3}$ (saturation density), $B_0=-16.0$~MeV (binding energy), 
$M^*_0/M_{\mbox{\tiny nuc}}=0.60$ (ratio of the effective mass to the nucleon rest mass), 
$K_0=230$~MeV (incompressibility), $J=31.6$~MeV (symmetry energy) and $L_0=58.9$~MeV, where 
$B_0=E(\rho_0) - M$, $M_0^*=M^*(\rho_0)$, $K_0=9(\partial p/\partial\rho)_{\rho_0}$, 
$J=\mathcal{S}(\rho_0)$, and $L_0=3\rho_0(\partial\mathcal{S}/\partial\rho)_{\rho_0}$, with 
$\mathcal{S}(\rho)=(1/8)(\partial^{2}E/\partial y^2)|_{y=1/2}$ and $E(\rho)=\epsilon/\rho$. In 
table~\ref{constants}, we show the values of the coupling constants of the model with and without 
the implementation of the SRC.
\begin{table}[!htb]
\centering
\caption{Coupling constants of the model determined by fixing some bulk parameters (see text). 
Results with and without inclusion of SRC. The remaining constant is fixed as  $C=0.005$ for both 
descriptions.}
\resizebox{\columnwidth}{!}{%
\begin{tabular}{ccc}
\hline
Coupling constant & Value (with SRC) & Value (without SRC) \\
\hline
$g_\sigma$  & $10.5550$   &  $10.4301$ \\
$g_\omega$  & $12.3443$   &  $13.5478$ \\
$g_\rho$    & $15.4069$   &  $11.1935$ \\
$A/M_{\mbox{\tiny nuc}}$       & $2.96396$   &  $1.77118$ \\
$B$         & $-29.8233$  &  $3.92522$ \\
$\alpha'_3$ & $0.00800924$ &  $0.0631432$ \\
\hline
\end{tabular}
}
\label{constants}
\end{table}

\begin{figure}[!htb]
\centering
\includegraphics[scale=0.3]{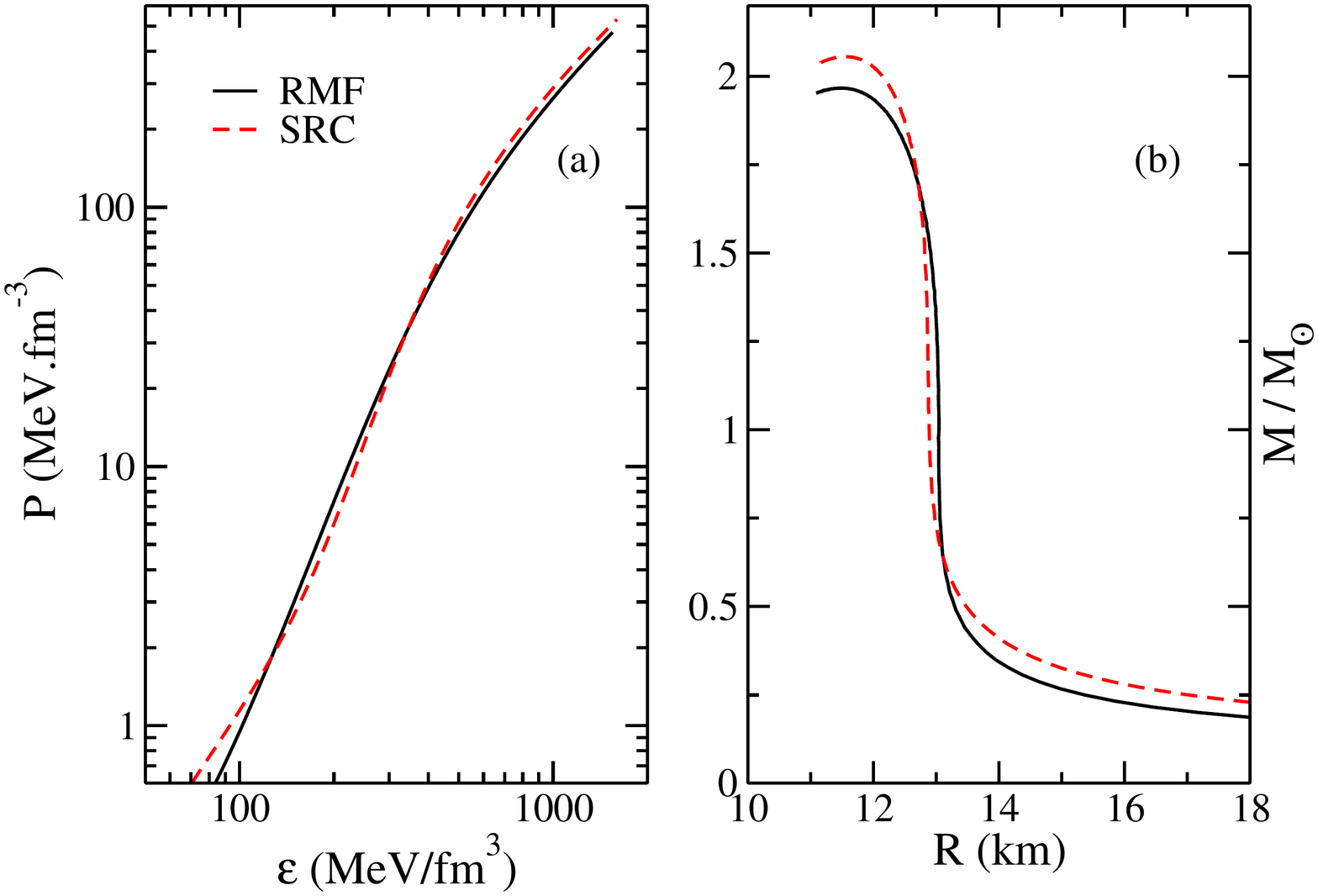}
\caption{Stellar matter constructed from the relativistic model with (SRC) and without (RMF) the 
inclusion of short-range correlations. (a) Total pressure as a function of total energy density. 
(b) Neutron star mass-radius diagram.}
\label{eos-src}
\end{figure}

In order to study the neutron star cooling it is necessary to impose charge neutrality and 
$\beta$-equilibrium to the system since we take into account the weak process and its inverse 
reaction, namely, $n\rightarrow p + e^- + \bar{\nu}_e$ and $p+ e^-\rightarrow n + \nu_e$, 
respectively. Here we consider stellar matter composed by protons, neutrons, electrons and muons, 
with the last particles emerging at densities for which the electron chemical potential ($\mu_e$) 
exceeds the muon mass ($m_\mu=105.7$~MeV). In our approach, neutrinos are assumed to escape due to 
their small cross-sections. This leads to the following conditions upon chemical potentials and 
densities: $\mu_n - \mu_p = \mu_e=\mu_\mu$ and $\rho_p - \rho_e = \rho_\mu$, where 
\mbox{$\rho_\mu=[(\mu_\mu^2 - m_\mu^2)^{3/2}]/(3\pi^2)$} and $\mu_e=(3\pi^2\rho_e)^{1/3}$. The total 
energy density and pressure of $\beta$-equilibrated stellar matter is then given by $\mathcal{E} = 
\epsilon + \epsilon_e + \epsilon_\mu$ and $P = p + p_e + p_\mu$, respectively. This specific EOS is 
show in Fig.~\ref{eos-src}{\color{blue}a}. Some neutron star properties, such as its mass-radius 
profile, can be found by solving the Tolman-Oppenheimer-Volkoff~(TOV) equations~\cite{tov39,tov39a}. 
Such a diagram is displayed in Fig.~\ref{eos-src}{\color{blue}b}.

The bulk parameters and their values used here to calibrate the coupling constants are the same 
studied in Ref.~\cite{Cai2016}. However, the output values of our couplings are different from those 
of Ref.~\cite{Cai2016} because we choose to fix $C=0.005$ instead of $C=0.01$. In 
Ref.~\cite{Cai2016}, the choice of the latter number implies a maximum neutron star mass of 
$M_{\mbox{\tiny{max}}}=1.87M_\odot$ and  $M_{\mbox{\tiny{max}}}=1.74M_\odot$, respectively, with and 
without the inclusion of SRC. These numbers do not satisfy the limits related to the PSR 
J1614-2230~\cite{dem10} and PSR J0348+0432~\cite{anto13} pulsars, namely,  $(1.97\pm 0.04)M_\odot$ 
and $ (2.01\pm 0.04)M_\odot$, respectively, neither the more recent prediction of 
$2.14^{+0.20}_{-0.18} M_\odot$ ($2.14^{+0.10}_{-0.09} M_\odot$) at 95.4\% (68.3\%) credible level 
from the MSP J0740+6620 pulsar~\cite{cromartie}. For this reason we choose to redefine the $C$ 
parameter in order to make the model compatible with these predictions. Since it is know that the 
decreasing of this parameter leads to a maximum neutron star mass increasing~\cite{serot}, we use 
here $C=0.005$. This modification implies in a model presenting $M_{\mbox{\tiny{max}}}=2.04M_\odot$ 
and $M_{\mbox{\tiny{max}}}=1.96M_\odot$ with and without SRC effects included, respectively. Notice 
that the effect of the SRC is to increase even further the maximum mass, as pointed out in 
Ref.~\cite{Cai2016}. The change in the $C$ parameter does not modify this feature.

\section{Neutron star cooling formalism}
We now investigate how the SRC influence the thermal evolution of neutron stars. We recall that 
the thermal evolution of a compact star is given by the general relativistic equations for thermal 
energy balance and transport, that can be written as 
\cite{2006NuPhA.777..497P,1999Weber..book,1996NuPhA.605..531S}
\begin{eqnarray}
  \frac{ \partial (l e^{2\phi})}{\partial m}& = 
  &-\frac{1}{\mathcal{E} \sqrt{1 - 2m/r}} \left( \epsilon_\nu 
    e^{2\phi} + c_v \frac{\partial (T e^\phi) }{\partial t} \right) \, , 
  \label{coeq1}  \\
  \frac{\partial (T e^\phi)}{\partial m} &=& - 
  \frac{(l e^{\phi})}{16 \pi^2 r^4 \kappa \mathcal{E} \sqrt{1 - 2m/r}} 
  \label{coeq2} 
  \, .
\end{eqnarray}

In order to solve Eqs.~(\ref{coeq1})-(\ref{coeq2}) one needs to combine information from the micro 
and macroscopic realms. The microscopic model provides information about the density/Fermi 
momentum/effective mass of all particles within the star which in turn can be used to calculate the 
specific heat, thermal conductivity and neutrino luminosity. The macroscopic information, obtained 
from solving the TOV equations furnish us with mass, radius and curvature information, as well as 
the spatial distribution of the aforementioned microscopic properties. One also needs boundary 
conditions, which is given by the  vanishing luminosity at the star center (which is evident as at 
$r = 0$ there is no heat flow) and by the thermal properties of the star atmosphere 
\citep{Gudmundsson1982,Gudmundsson1983,Page2006}. In this work we consider the simplest atmosphere 
possible, without strong magnetic field and without accreted matter due to fallback. 

With all micro and macroscopic data, as well as the proper boundary conditions we perform numerical 
calculations to obtain detailed solution to Eqs.~(\ref{coeq1})-(\ref{coeq2}) for neutron stars 
covering a wide range of masses in  both the RMF and SRC models. We note that all possible neutrino 
emissivities processes are accounted for, most prominent of which is the direct Urca (DU) process, 
modified Urca process (MU) and Bremsstrahlung (BR) process (all of which takes place in the star 
core). For a detailed review of such processes we refer the reader to 
\cite{Yakovlev2000,Yakovlev2004}. 

Particular attention must be given to the DU process  - as it is significantly more powerful than 
other neutrino emission processes. The DU process consists of the neutron beta decay ($n \rightarrow 
p ~+~ e^{-} + \bar{\nu} $) and proton electron capture ($p~ +~  e^{-} \rightarrow n + \nu$). It has 
a very large luminosity ($\sim 10^{27} (T_9)^6$ erg/cm$^3$s), although it can only take place if the 
triangle inequality $k_{fn} \leqslant k_{fp} + k_{fe}$ is satisfied. This generally mean that it 
will only occur for high enough proton fractions (usually $\sim 11 - 15\%$ 
\cite{Lattimer1991,Page2006}). This is particularly important for the work presented in this paper, 
as the SRC will substantially modify the proton Fermi momentum, which subsequently modify the proton 
fraction within the star. Thus, as we will discuss below, the inclusion of the SRC can lead to 
substantially different cooling properties, even though the overall macroscopic properties of the 
star (mass, radius) and bulk properties of the model are not too different.

\section{Results}
\begin{figure}[!htb]
\centering
\includegraphics[scale=0.3]{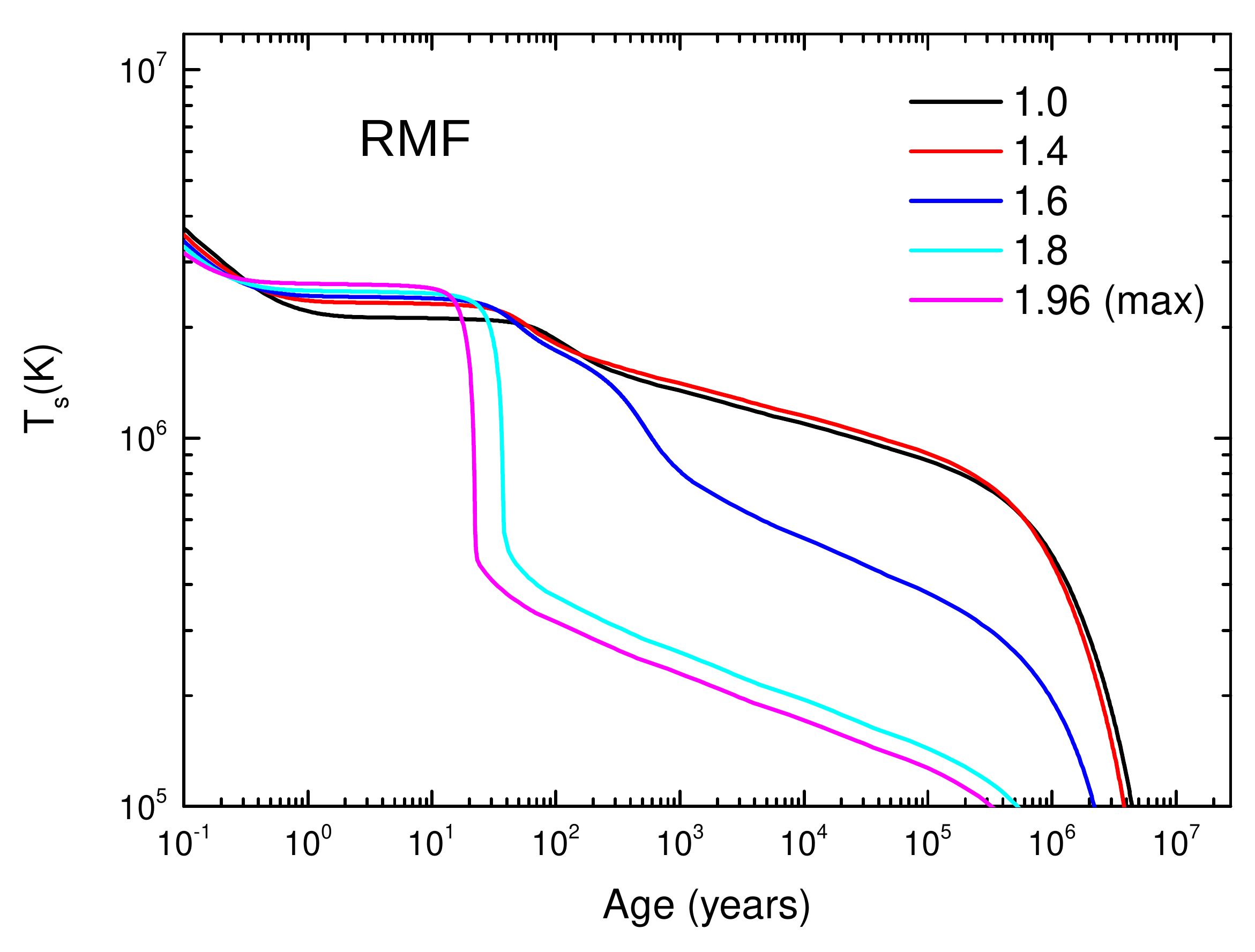}
\caption{Redshifted surface temperature evolution for the RMF model. The labels indicate the masses 
of the star in solar masses}
\label{RMFcool}
\end{figure}

\begin{figure}[!htb]
\centering
\includegraphics[scale=0.3]{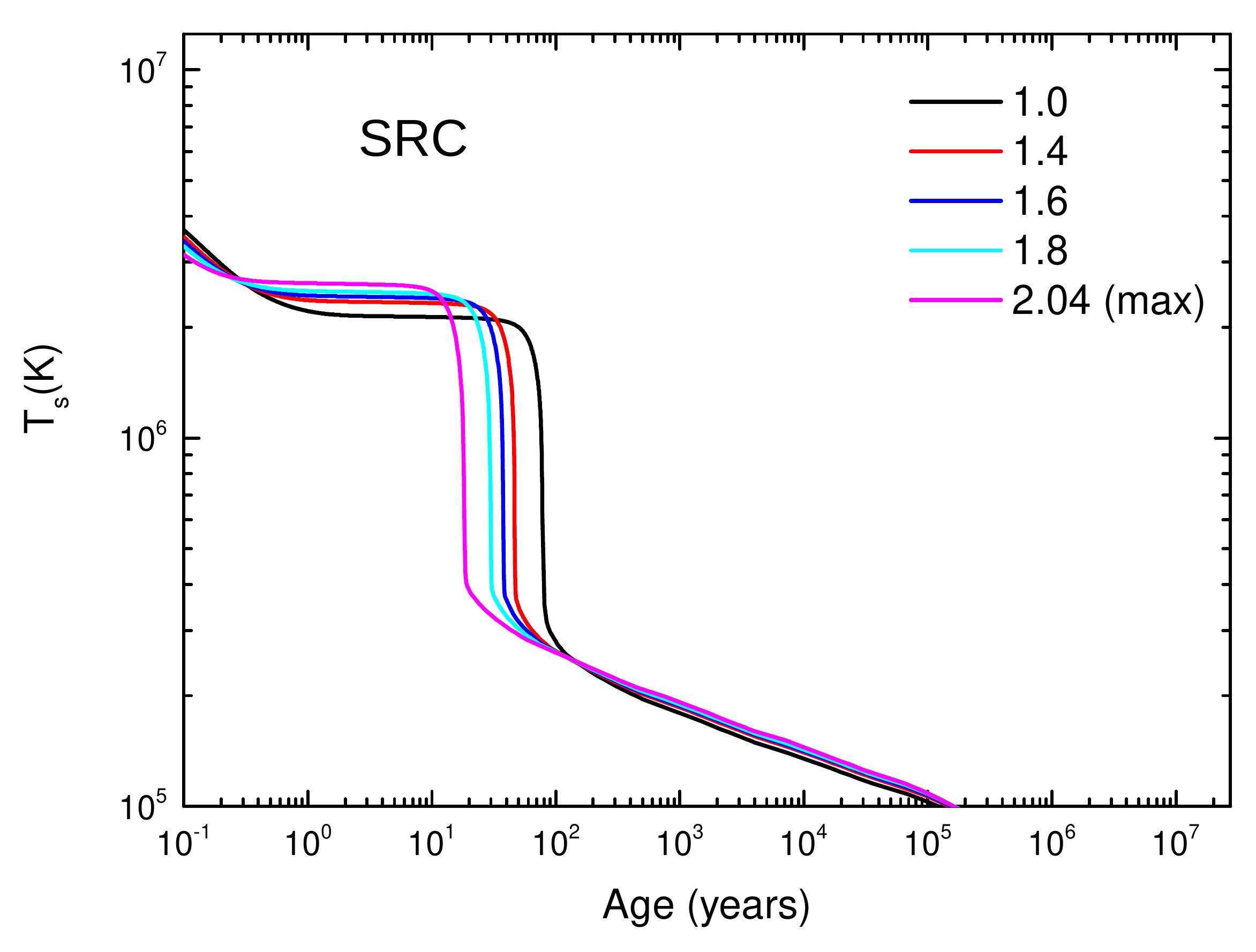}
\caption{Same as Fig.~\ref{RMFcool} but for the short-range-correlation (SRC) model.}
\label{SRCcool}
\end{figure}

We now show the thermal evolution of stars described by the microscopic models discussed in the 
previous sections. We show the cooling of stars with different masses, representative of the 
families shown in Fig.~\ref{eos-src}. The thermal evolutions of such stars are plotted in 
Figs.~\ref{RMFcool} - \ref{SRCcool}
\textbf{}

The results depicted in Figs.~\ref{RMFcool} - \ref{SRCcool} clearly show a significant difference 
in the qualitative cooling behavior of such stars, which is somewhat surprising given that both 
models have similar bulk properties, namely, $\rho_0=0.15$~fm$^{-3}$, $B_0=-16.0$~MeV, 
$M^*_0/M_{\mbox{\tiny nuc}}=0.60$, $K_0=230$~MeV, $J=31.6$~MeV and $L_0=58.9$~MeV. This indicates 
that the mere presence of short-range-correlations is enough to cause a significant deviation in the 
cooling nature of the star (especially for more massive stars). We can see that in the absence of 
short range correlations (RMF model) stars with masses up to $\sim 1.6 M_\odot$ exhibit slow 
cooling. Stars above this mass start to allow for the DU process, leading to a fast cooling 
behavior. The inclusion of short-range-correlations (SRC model) drastically change this, and all 
stars in the model allow the DU process. 

\begin{figure}[!htb]
\centering
\includegraphics[scale=0.3]{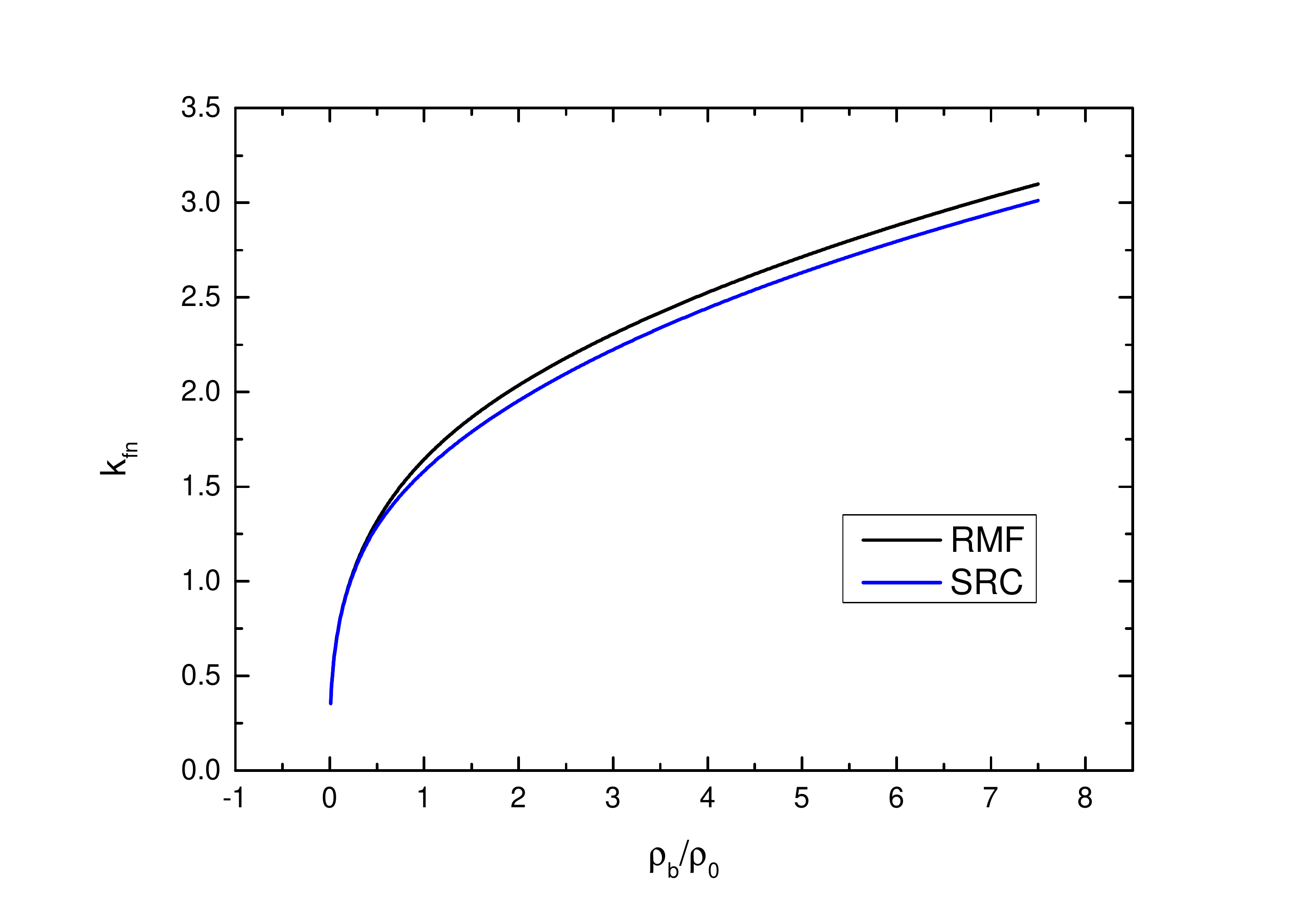}
\caption{Neutron Fermi momentum as a function of baryon density for the model with (SRC) and 
without short-range-correlations.}
\label{kfn}
\end{figure}

\begin{figure}[!htb]
\centering
\includegraphics[scale=0.3]{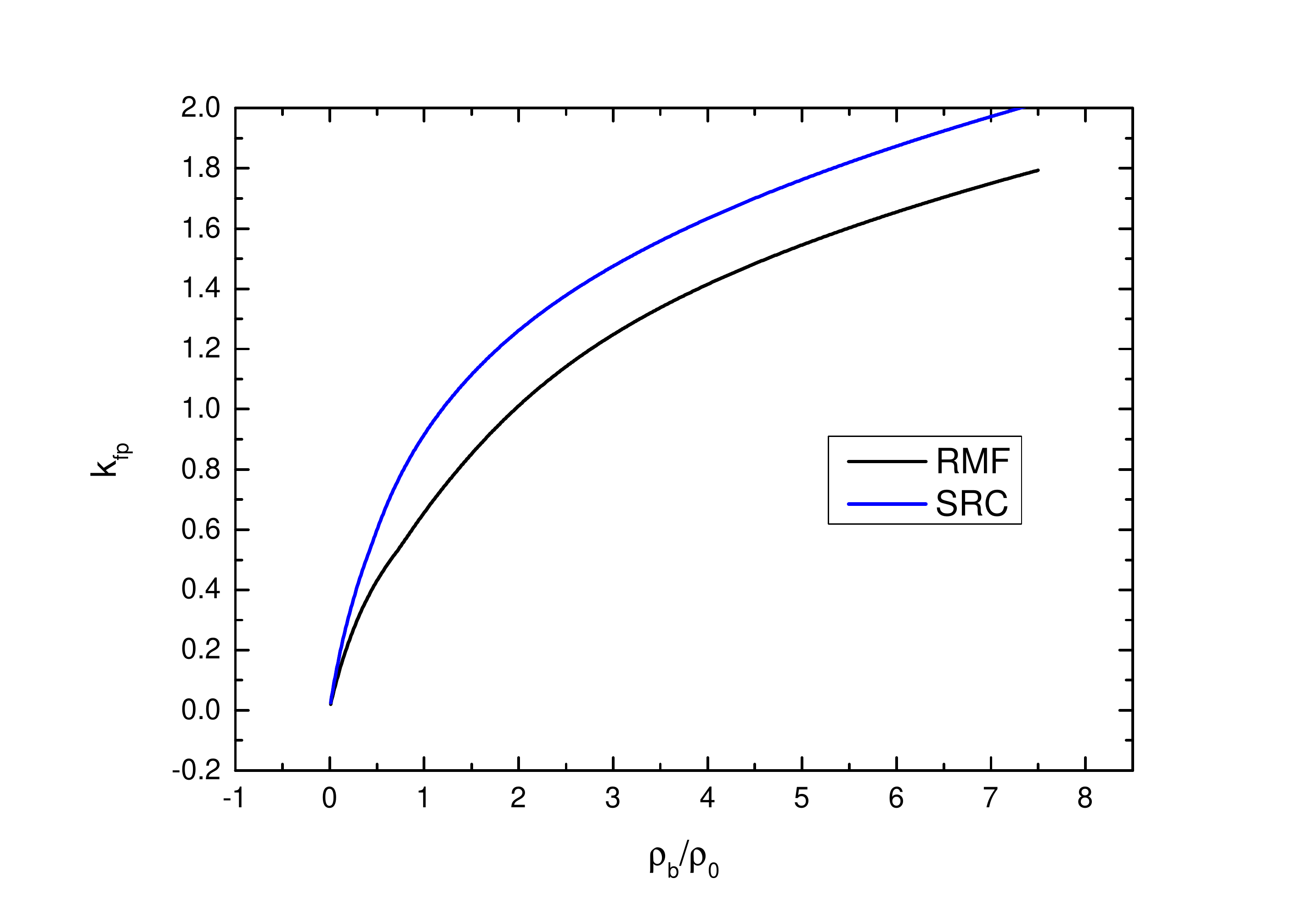}
\caption{Same as Fig.~\ref{kfn} but for the proton Fermi momentum.}
\label{kfp}
\end{figure}

The behavior exhibited by stars with SRC indicates that the threshold above which the DU is allowed 
is drastically reduced by the presence of SRC. This indicates that Fermi momentum of the particles 
that take place in the DU process should be drastically different between the RMF and SRC models. 
This is confirmed by Figs.~\ref{kfn}-\ref{kfp} where we show the neutron and proton Fermi momentum 
as a function of density for the two models studied.

\begin{figure}[!htb]
\centering
\includegraphics[scale=0.3]{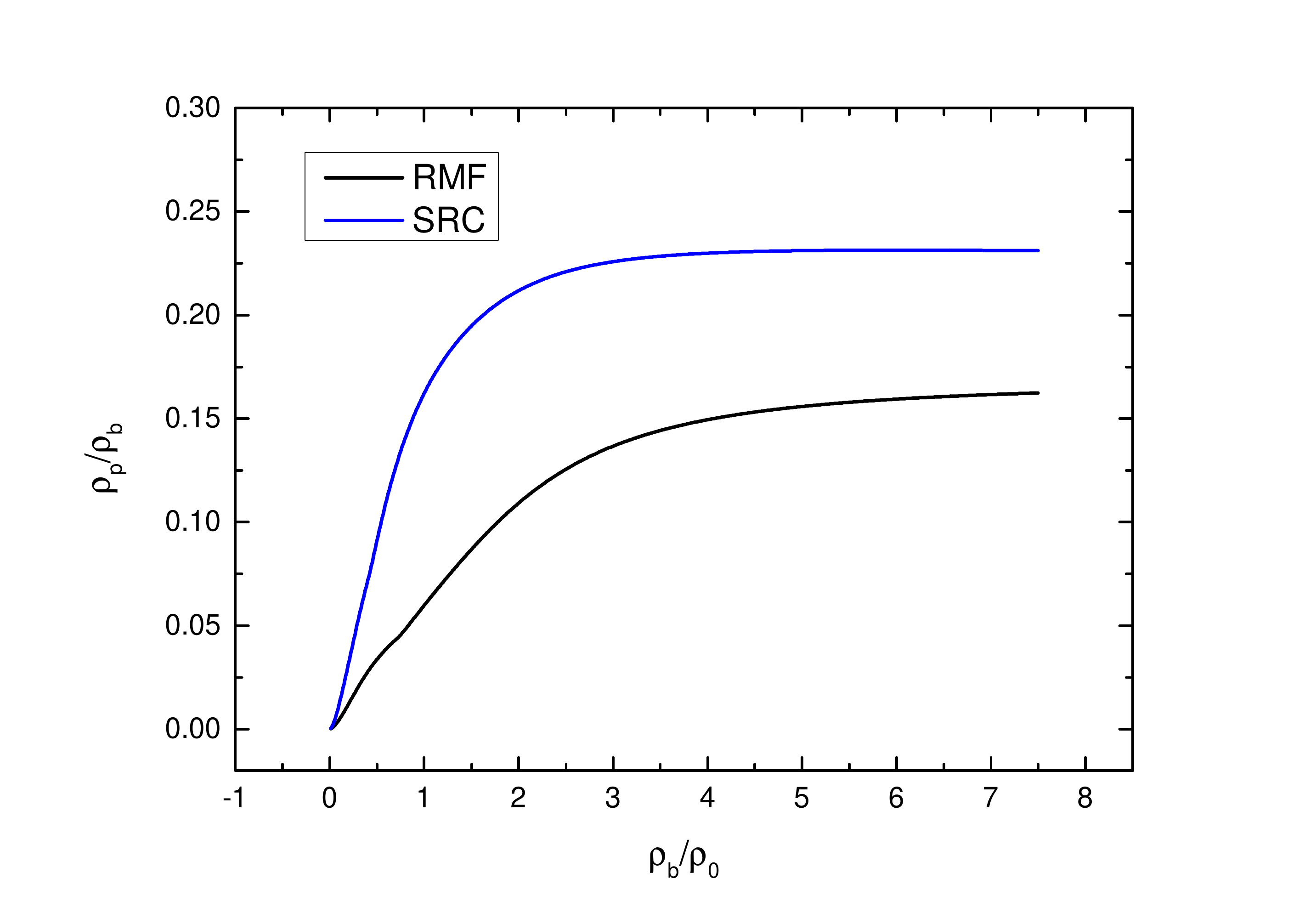}
\caption{Proton fraction as a function of relative baryon density for the model with 
short-range-correlations (SRC) and without (RMF).}
\label{pfrac}
\end{figure}

While Fig.~\ref{kfn} shows us that (at the same density) the neutron Fermi momentum is slightly 
smaller for the model with SRC, there is an equivalent increase in the proton Fermi momentum. Such 
interchange leaves the baryon density unchanged but significantly increases the proton fraction for 
the SRC model, thus leading to an early onset of the DU process. This is confirmed by 
Fig.~\ref{pfrac} where we see the proton fraction as a function of density - which shows a 
significantly higher proton fraction for the SRC model. 

\section{Summary and concluding remarks}

In this work we investigated the influence of SRC on the cooling of neutron stars. For that 
purpose, we based our study on a relativistic mean field model with SRC phenomenology firstly 
implemented in Ref.~\cite{Cai2016}. Basically, the modification in the RMF model comes from the 
momentum distribution function, $n(k)$, that now takes into account a high momentum tail in which 
$n(k)$ depends on $k$ as $n(k) \sim k^{-4}$ for values greater than the Fermi momentum. The change 
in $n(k)$ results in the split of the momentum integrals present in the energy density, pressure 
and scalar density of the model into two parts, as one can see in Eqs.~(\ref{rhos}) to 
(\ref{press}). As in Ref.~\cite{Cai2016}, we also fixed the bulk parameters of the model in the same 
values, as already mentioned. However, we fixed the coupling constant $C$ to the value of $0.005$ 
instead of $C=0.01$ previously used in Ref.~\cite{Cai2016} because this new value ensures the model 
presents $M_{\mbox{\tiny{max}}}=2.04M_\odot$ and $M_{\mbox{\tiny{max}}}=1.96M_\odot$ with and 
without SRC effects, respectively (see Fig.~\ref{eos-src}), in agreement with the limits of recently 
detected pulsars.

This study has demonstrated that short range correlations play an important role in the thermal 
evolution of neutron stars. Our investigation has shown that for two equivalent models, the one with 
the inclusion of SRC exhibits a drastically different cooling behavior. We have tracked the origin 
of such difference back to the proton and neutron Fermi momenta and have shown that for the SRC 
model there is an increase in the proton Fermi momentum, accompanied by a reduction of the neutron 
one. This leads to approximately the same baryon density but with a significantly larger proton 
fraction for the SRC model - which in turn leads to an early onset of the DU process. This is 
reflected in the thermal evolution of SRC stars, which undergo a much faster cooling than their 
non-SRC counterparts. We conclude that SRC potentially influence the thermal evolution of neutron 
stars, although they do not alter considerably the macroscopic properties of such objects. As we 
have shown in this study, the EoS and consequently the Mass-Radius diagrams of both the RMF and SRC 
are largely similar, and yet there is a significant difference in the thermal evolution of the 
family of stars these EOS describe. Further observation of neutron star cooling may potentially shed 
some light on the intensity of SRC in neutron star matter, or equivalently further understanding of 
SRC may aid us in improving our comprehension of the cooling of neutron stars.

\section{Acknowledgments}
This work is a part of the project INCT-FNA Proc. No. 464898/2014-5, partially supported by 
Conselho Nacional de Desenvolvimento Cient\'ifico e Tecnol\'ogico (CNPq) under grants No. 
310242/2017-7 and 406958/2018-1 (O.L.), 433369/2018-3 (M.D.), 301155/2017-8 (D.P.M.), and by 
Funda\c{c}\~ao de Amparo \`a Pesquisa do Estado de S\~ao Paulo (FAPESP) under the thematic projects 
2013/26258-4 (O.L.), 2019/07767-1 (L.A.S.) and 2017/05660-0 (O.L., M.D.). L.A.S. thanks the support 
by INCT-FNA Proc. No. 88887.464764/2019-00.

\bibliographystyle{elsarticle-num1}
\bibliography{bib}

\end{document}